\documentclass[twocolumn,secnumarabic,amssymb,nobibnotes,aps,pra,showpacs,superscriptaddress]{revtex4-1}
\usepackage[english]{babel}
\usepackage{graphicx}
\usepackage{amsmath}
\usepackage{amsfonts}
\usepackage[usenames,dvipsnames]{color}
\usepackage{hyperref}
\usepackage{blindtext}
\hypersetup{colorlinks}
\hypersetup{linkcolor=black, citecolor=black, urlcolor=blue}
\usepackage[utf8]{inputenc}
\usepackage{color}

\begin{document}
\title{Charge-, salt- and flexoelectricity-driven anchoring effects in nematics}
\author{Jeffrey C. Everts}
\email{jeffrey.everts@gmail.com}

\address{Faculty of Mathematics and Physics, University of Ljubljana, Jadranska 19, 1000 Ljubljana, Slovenia}

\author{Miha Ravnik}
\address{Faculty of Mathematics and Physics, University of Ljubljana, Jadranska 19, 1000 Ljubljana, Slovenia}
\address{Jozef Stefan Institute, Jamova 39, 1000
Ljubljana, Slovenia}

\date{\today}

\begin{abstract}
We investigate the effects of electric double layers and flexoelectricity on the surface anchoring in general nematic fluids. Within a simplified model, we demonstrate for a nematic electrolyte how the surface anchoring strength can be affected by the surface charge, bulk ion concentration and/or flexoelectricity, effectively changing not only the magnitude of the anchoring but also the anchoring type, such as from planar to tilted. In particular, we envisage possible tuning of the anchoring strength by the salt concentration in the regime where sufficiently strong electrostatic anchoring, as controlled by the (screened) surface charge, can compete with the non-electrostatic anchoring. This effect is driven by the competing energetic-torque couplings between nematic director and the emergent electrostatic potential, due to surface charge, ions and flexoelectricity. Our findings propose a way of influencing surface anchoring by using electrostatic effects, which could be used in various aspects, including in the self-assembly of colloidal particles in nematic fluids, optical and display patterns, and sensing.
\end{abstract}

\maketitle

\section{Introduction}
Surfaces are one of the main tools for controlling the orientational order of the building blocks in complex nematic fluids. The interaction imposed at the surfaces on the nematic fluids is called surface anchoring \cite{Jerome:1991, deGennes:1993}, and it fundamentally emerges because of the interaction between the nematic building blocks and the surface material,  such as a polymer-rubbed wall, glass or another liquid or gas. With span of combinations of materials, the surface anchoring can impose surface orientational order in almost every arbitrary direction, from homeotropic \cite{Stark:2001}, planar \cite{Crawford:1993}, planar degenerate \cite{Fournier:2005} to tilted \cite{Patel:1993} and even degenerate tilted \cite{Jagemalm:1997}; also the interaction can be of different strength. Technologically, control over surface anchoring is instrumental in display applications of nematic fluids, but goes also far beyond this, and is central in applications including sensors \cite{Crawford:2007}, emulsification \cite{Poulin:1998} and microfluidics \cite{Batista:2015}. The surface anchoring is commonly associated with surface-imposed ordering of (passive) materials such as molecules, polymers, or colloidal particles, but actually, the same type of interaction emerges also in non-equilibrium -active matter- systems \cite{Wu:2017,Duclos:2018}.   

An electric field is another tool for controlling the orientational order in nematic fluids, and is usually applied as an external field (voltage), but can also be sourced, for example, from local charges present in the materials (such as ions) \cite{Sawada:1999}, or by flexoelectricity where the elastic distortion of the nematic generates a local (electric) polarisation \cite{Meyer:1969}. Often the effects of free charges and flexoelectricity are small and considered negligible, but this does not need to be true. By introducing ions into nematics or using flexoelectricity, one opens a possibility to create electric fields that are strongly spatially anisotropic both in direction and magnitude, with the complexity of the electric field being determined by the nematic and ion profile (and their mutual coupling), and not only by the electrodes as in the case of external electric fields \cite{Thurston:1984}. Furthermore, the use of ions in liquid crystals was demonstrated in a recent experimental paper to affect electrostatically controlled anchoring of charged colloidal particles on a surface \cite{Mundoor:2019}, and also indirectly by the balance of screened electrostatic interactions and elastic interactions to produce specific crystal structures \cite{Mundoor:2016}. Ions also influence the isotropic-nematic phase transition in lyotropic liquid crystals \cite{Giesselmann:2005} and influences nonlinear electroosmosis in nematic liquid crystals \cite{Poddar:2017}. On the other hand, flexoelectricity is predicted to stabilize blue plases \cite{Alexander:2007, Coles:2010} and is known to influence director profiles in nematic cells \cite{Durand:1986, Zumer:1999, Liu:2018}, colloidal transport \cite{Terentjev:2007}, and flow \cite{Toth:2005, Pieranski:2019}.
The joint phenomenon in these works is that the anisotropic dielectric interaction of  electric charge or flexoelectric polarisation in a nematic directly competes with other aligning mechanisms of nematics, such as the nematic elasticity and the surface anchoring.

In this paper, we show the possible control and tuning of nematic surface ordering, beyond standard surface anchoring, using surface charge, bulk ionic inclusions and flexoelectricity, as also affected by charge regulation. Using a simplified model, we obtain a better general understanding of how the surface charge, electrostatic screening by ions and flexoelectricity in combination with standard surface anchoring can lead to various effective surface anchoring regimes, as a result of various combined local torques, which are not only of different effective strengths, but also types. This work extends previous work on salt-induced surface anchoring \cite{Petrov:1993, Lavrentovich:1994, Lev:1999, Meister:1999, Schmiedel:1999, Ziherl:1999, Barbero:2000, Shah:2001, Derfel:2002, Barbero:2002, Evangelista:2006}, by putting more emphasis on varying the salt concentration by e.g. doping the sample, which affects the anchoring via electrostatic screening and charge regulation, as was already experimentally observed in Ref. \cite{Bungabong:2010}. Finally, the work is a contribution towards establishing electrostatic effects -charge, ions, flexoelectrictiy- as an interesting control mechanism for (self) assembly and even topology in general nematic complex soft matter.

This paper is organised as follows. In Sec. \ref{sec:theory} we outline the theoretical framework that we will use. In Sec. \ref{sec:chargean} we discuss the control of anchoring by surface charge and salt, and discuss the influence of the various parameters, while in Sec. \ref{sec:CR} we highlight the effect of charge regulation. In Sec. \ref{sec:flexo}, we show how flexoelectricity influences the anchoring and in Sec. \ref{sec:con} we finish with conclusions and an outlook.

\section{Theory}
\label{sec:theory}
\begin{figure}[t]
\centering
\includegraphics[width=0.45\textwidth]{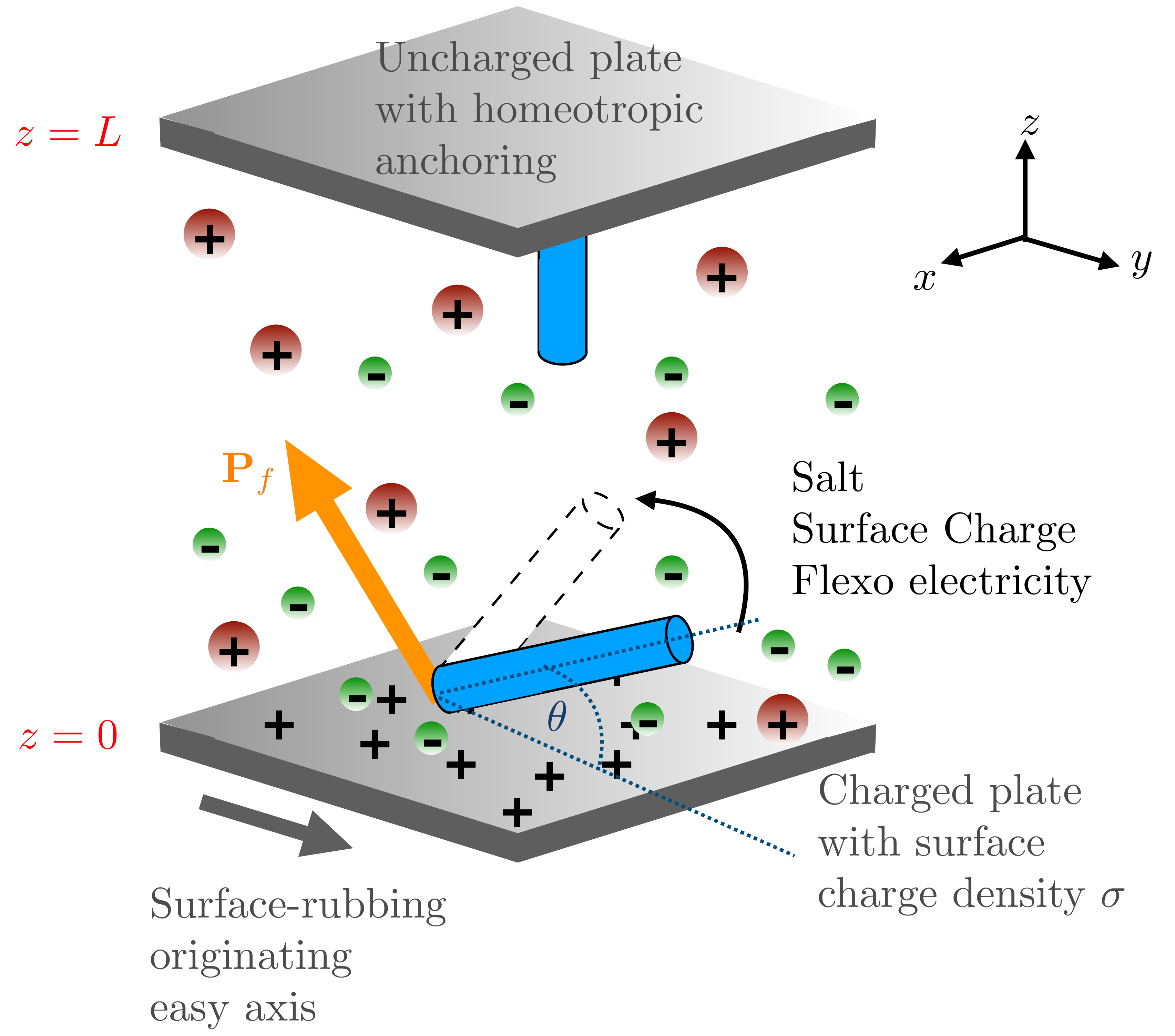}
\caption{Scheme of the regime for charge and flexoelectricity controlled effective surface anchoring. Between a charged plate with surface charge density $\sigma$ at $z=0$ and an uncharged plate at $z=L$, positive and negative ions are dispersed in a nematic medium. The bottom plate is assumed to have  standard weak-anchoring conditions with easy axis equal to the $y$ axis, whereas the upper plate has strong homeotropic boundary conditions for the liquid crystal. Locally in the whole region between the plates, there is a flexoelectric polarisation ${\bf P}_f(z)$ and director angle $\theta(z)$; blue cylinders indicate the local director.}
\label{fig:schematic}
\end{figure}

The design of effective surface anchoring is  demonstrated in the geometry of standard (hybrid aligned nematic - HAN) nematic cells, as shown in Fig.~\ref{fig:schematic}. This simple geometry is chosen for clearness of the effects, but the demonstrated phenomena could be applied in principle in any nematic cell geometry, nematic emulsions, or nematic colloids, performing as a major novel control mechanism in these systems. Specifically, we consider a single charged plate with charge density $q_e\sigma$ at $z=0$, separated from an uncharged plate situated at $z=L$, both with surface area $A$. Between the two plates, a nematic material is present which is characterised by the dielectric tensor $\epsilon_{ij}({\bf r})=\epsilon_\perp\delta_{ij}+\Delta\epsilon{n}_{i}({\bf r}){n}_j({\bf r})$, where $\epsilon_\perp$ is the dielectric constant perpendicular to the director field ${\bf n}({\bf r})$, and $\Delta\epsilon$ is the dielectric anisotropy. For the upper plate we assume strong homeotropic anchoring conditions, while for the bottom plate weak planar non-degenerate anchoring is assumed with the easy axis parallel to the plate. This cell geometry allows us to parametrise the nematic director as ${\bf n}({\bf r})=(0, \cos[\theta(z)], \sin[\theta(z)])$. 

The nematic fluid with dispersed ions and emergent flexoelectricity can be described at the continuum mesoscopic level by a total free energy $\mathcal{F}[\phi,\rho_\pm,\theta]$ which is a combination of various free energy contributions: nematic elasticity, surface anchoring, (ideal-gas) entropy of ions in nematic, and the electrostatic energy.  
Therefore, the free energy is written as,
\begin{equation}
\mathcal{F}[\phi,\rho_\pm,\theta]=\mathcal{F}_\text{LC}[\theta]+\mathcal{F}_\text{S}[\rho_\pm]+\mathcal{F}_\text{EL}[\phi,\rho_\pm,\theta],
\label{fe}
\end{equation}
where the liquid-crystal part is described by the Oseen-Frank elastic free energy \cite{Oseen:1933, Frank:1958} within the equal-constant approximation and the anchoring is described by the Rapini-Papoular surface free energy \cite{Rapini:1969},
\begin{equation}
\frac{\mathcal{F}_\text{LC}[\theta]}{A}=\frac{1}{2}\int_0^L dz\, K[\theta'(z)]^2-\frac{W}{2}\cos^2[\theta(z=0)],
\end{equation}
where $K$ is the elastic constant of the nematic, $W$ is the anchoring strength and the prime $'$ indicates differentiation with respect to $z$. In the nematic, ions are dissolved with number density $\rho_\pm(z)$, and they give an (ideal-gas) contribution to the free energy \cite{Barrat:2003},
\begin{equation}
\frac{\beta\mathcal{F}_\text{S}[\rho_\pm]}{A}=\sum_\alpha\int_0^Ldz\, \rho_\alpha(z)\{\ln[\rho_\alpha(z)\Lambda_\alpha^3]-1\}
\label{eq:electricenergy}
\end{equation}
where $\beta^{-1}=k_BT$ is the thermal energy and $\Lambda_\pm^3$ the thermal volume. The electrostatic part couples $\rho_\pm(z)$ and $\theta(z)$, and we will express this with the formulation of the free energy by using the electrostatic potential $\phi(z)/(\beta q_e)$ as a variational parameter,
\begin{align}
\frac{\beta\mathcal{F}_\text{EL}[\phi,\rho_\pm,\theta]}{A}=\int_0^L dz\, \Bigg\{q(z)\phi(z)+\phi'(z) (P_f)_z(z) \nonumber \\
-\frac{1}{8\pi\lambda_B}\tilde\epsilon_{zz}(z)[\phi'(z)]^2 \Bigg\}.
\end{align}
Here, $q(z)=\rho_+(z)-\rho_-(z)+\sigma\delta(z)$ is the total (number) charge density, we rescaled the dielectric tensor $\tilde{\epsilon}_{ij}(z)=\epsilon_{ij}(z)/\bar{\epsilon}$ using the rotationally averaged dielectric constant, $\bar{\epsilon}=(\mathrm{Tr}\boldsymbol{\epsilon})/3$, and $\lambda_B=q_e^2/(4\pi\epsilon_0\bar{\epsilon}k_BT)$ is the isotropic Bjerrum length. In this geometry $\epsilon_{zz}(z)=\epsilon_\perp+\Delta\epsilon\sin^2[\theta(z)]$.

The flexoelectric polarisation $q_e{\bf P}_f(z)$ is given by \cite{Barbero:1986}
\begin{equation}
{\bf P}_f=e_1{\bf n}(\nabla\cdot{\bf n})-e_3[{\bf n}\times(\nabla\times{\bf n})],
\end{equation}
with $q_ee_1$ and $q_ee_3$ the splay and bend flexoelectric coefficient, and $q_e$ the elementary charge, such that $-\nabla\cdot{\bf P}_f$ is a number (bound) charge density with same dimensions as $\rho_\pm(z)$. To convert to charge densities given in [C m$^{-3}$] one should simply multiply with $q_e$. In this particular geometry, the flexoelectric polarisation is of the form
\begin{equation}
{\bf P}_f=\left(0,e_1\cos^2\theta-e_3\sin^2\theta,\frac{e_1+e_3}{2}\sin(2\theta)\right)\theta'(z).
\end{equation}

The equilibrium profile of the nematic is determined by the minimum of the total free-energy (Eq.~\ref{fe}), which is minimised according to electrostatic potential $\phi(z)$, nematic orientation (given by director angle $\theta(z)$), and ion number densities $\rho_\pm(z)$. Firstly, minimisation with respect to the electrostatic potential $\phi(z)$, $\delta\mathcal{F}/\delta{\phi}(z)=0$, gives the Poisson equation,
\begin{equation}
\left[\tilde{\epsilon}_{zz}(z)\phi'(z)-4\pi\lambda_B(P_f)_z(z)\right]'=-4\pi\lambda_Bq(z),
\label{eq:poisson}
\end{equation}
or rewritten in SI units $\nabla\cdot{\bf D}=q_eq({\bf r})$, with \mbox{${\bf D}=-\epsilon_0\boldsymbol{\epsilon}\cdot\nabla\phi/(\beta q_e)+q_e{\bf P}_f$} the dielectric displacement. Moreover, notice that $\min_\phi\mathcal{F}_\text{EL}[\phi,\rho_\pm,\theta]=(1/2)\int d{\bf r}\, q({\bf r})\phi({\bf r})$, which is the familiar electrostatic energy. This shows that we indeed used the correct (variational) free energy. The surface terms in the variation of the free energy with respect to the electrostatic potential give the electrostatic boundary conditions
\begin{align}
\tilde{\epsilon}_{zz}(z)\phi'(0)-4\pi\lambda_B(P_f)_z(0)=-4\pi\lambda_B\sigma,
\label{eq:chargeboundary}
\end{align}
and global charge neutrality condition
\begin{align}
\tilde{\epsilon}_{zz}(L)\phi'(L)-4\pi\lambda_B(P_f)_z(L)=0.
\label{eq:neutralboundary}
\end{align}
Secondly, minimisation of the total free energy with respect to the nematic orientation, $\delta\mathcal{F}/\delta\theta(z)=0$, gives the nematic Euler-Lagrange equation for the nematic director 
\begin{equation}
\beta K\theta''(z)+\left\{\frac{\Delta\tilde{\epsilon}}{8\pi\lambda_B}[\phi'(z)]^2+\frac{e}{2}\phi''(z)\right\}\sin[2\theta(z)]=0,
\label{eq:eultheta}
\end{equation}
with $e=e_1+e_3$ and boundary conditions for the nematic as
\begin{equation}
\theta'(0)=\frac{1}{2}\sin[2\theta(0)]\left[\frac{1}{\xi_s}+\frac{e}{\beta K}\phi'(0)\right],\ \theta(L)=\frac{\pi}{2},
\label{eq:anchor}
\end{equation}
where $\xi_s=K/W$ is the nematic surface extrapolation length \cite{deGennes:1993}. 

Thirdly, minimisation of the total free energy with respect to $\rho_\pm(z)$, $\delta\mathcal{F}/\delta\rho_\pm(z)=\mu_\pm$, gives the final Euler-Lagrange equation, where $\beta\mu_\pm=\log(\rho_s\Lambda_\pm^3)$ is the chemical potential of the reservoir (ions are treated grand canonically with fixed chemical potential, so total number of ions is \emph{not} fixed), with reservoir salt concentration $\rho_s$. We find the Boltzmann distribution
\begin{equation}
 \rho_\pm(z)=\rho_s\exp[\mp\phi(z)].
\end{equation}
Combined with Eq. \eqref{eq:poisson}, this introduces the modified Poisson-Boltzmann equation, that now full incorporates not only bulk ions but also flexoelectricity (and naturally, the coupling to nematic director profiles via the dielectric tensor),
\begin{equation}
\left[\tilde{\epsilon}_{zz}(z)\phi'(z)-4\pi\lambda_B(P_f)_z(z)\right]'=\kappa^2\sinh[\phi(z)],
\label{eq:PB}
\end{equation}
where we introduced the isotropic Debye screening length $\kappa^{-1}=\lambda_D =(8\pi\lambda_B\rho_s)^{-1/2}$. Eqs.  \eqref{eq:chargeboundary}-\eqref{eq:PB} form a closed set of equations which we solve numerically by using the finite-element software package COMSOL Multiphysics.

The following numerical parameters explained below will be used in numerical calculations, roughly corresponding to the standard liquid crystal 5CB at $T=298$K (which is roughly $10$K below the isotropic-nematic transition temperature). Dielectric permittivities are $\Delta\epsilon=13$, $\epsilon_\perp=6$, and the isotropic dielectric constant is $\bar{\epsilon}\approx10$ \cite{Bogi:2001}, which gives $\lambda_B=6\ \mathrm{nm}$. For the elastic constant we take the average between the splay and twist constant, $K=8\times10^{-12}$ N \cite{Bogi:2001} and the thermal energy is $k_BT=4\times10^{-21}$ J. Anchoring strengths are considered between $10^{-3}$ J m$^{-2}$ to $10^{-8}$ J m$^{-2}$, corresponding to strong-anchoring and weak-anchoring regimes, respectively \cite{Blinov:1989}, giving surface extrapolation lengths between $10^{-10}$ m and $10^{-5}$ m. 

The values of the surface charge density $\sigma$ in nematic solvents like 5CB are less well known. For silica in water $\sigma$ can vary between $10^{-3}$ nm$^{-2}$ and $10$ nm$^{-2}$; however, 5CB has a much smaller dielectric constant than water, so we expect lower surface charges. Similar dielectric (but isotropic) solvents reported surface charge densities between $10^{-5}$ nm$^{-2}$ and $10^{-4}$ nm$^{-2}$ for charged colloidal particles in cyclohexylbromide with dielectric constant $\epsilon=7.92$ \cite{Linden:2015, Everts:2016}, so we will use these as order of magnitude estimates for $\sigma$.  Finally, we vary the Debye screening length $\kappa^{-1}=\lambda_D$ between $10^{-9}$ m and $10^{-5}$ m, based on estimated Debye screening lengths due to ionic impurities in some liquid crystals \cite{Thurston:1984b, Shah:2001, Colpaert:1996}, or estimates of liquid crystals with doped ions \cite{Cognard:1984, Sprokel:1973, Sprokel:1974, Bungabong:2010}. The nanometre regime could be accessible if special salts are synthesised that can dissolve in low-dielectric liquid crystalline solvents, or if water-based liquid crystal systems are used such as lyotropic or chromonic liquid crystals (although these systems might not have a significant dielectric anisotropy). 

Finally, flexoelectric coefficients $q_e(e_1+e_3)$ of 5CB are estimated to be of the order 1-10 pC m$^{-1}$ \cite{Murthy:1993}, while also much higher values are currently obtained with special molecules, the so-called giant flexoelectricity with flexoelectric coefficient up to nC m$^{-1}$ \cite{Harden:2006}. Note that in this geometry only the sum $e=e_1+e_3$ matters, because the $y$ component of the flexoelectric polarisation is projected out because of translational symmetry in the plane of the surfaces. In this work we will use the flexoelectric coefficients as tuning parameters and use the pC m$^{-1}$ regime.

\section{Electric charge-controlled effective anchoring}
\label{sec:chargean}

\begin{figure}[t]
\centering
\includegraphics[width=0.47\textwidth]{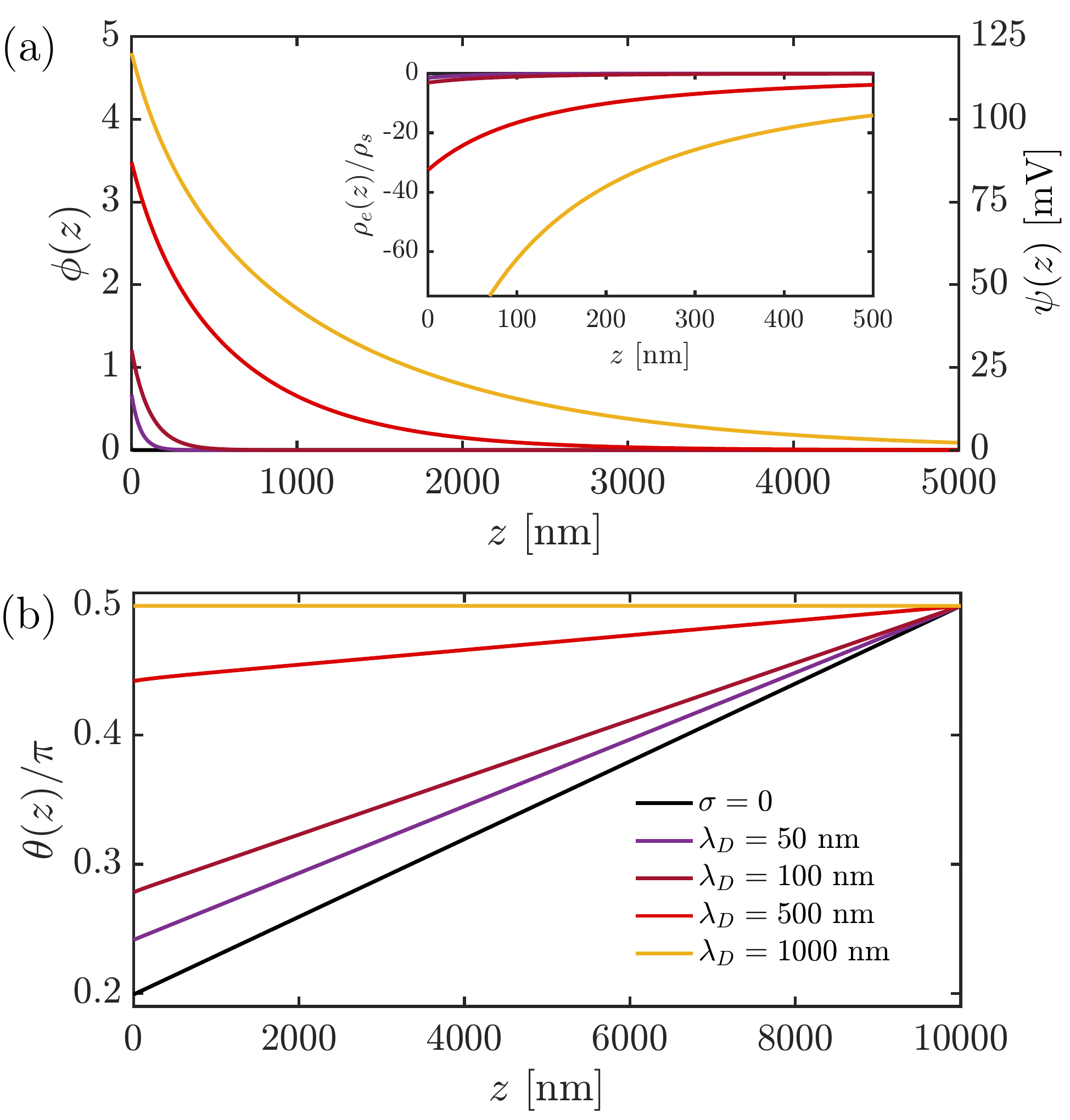}
\caption{Charge-controlled nematic director and electrostatic potential profiles in a hybrid aligned nematic (HAN) cell. (a) Electrostatic potential $\psi(z)=\phi(z)/(\beta q_e)$ profiles and (b) nematic director profiles $\theta(z)$ along the cell thickness $z$, for different isotropic Debye screening lengths $\lambda_D$. Inset in (a) shows the net charge number density in the diffuse screening cloud $\rho_e(z)$, normalised to the bulk concentration $\rho_s$. In all calculations we set $\sigma=2\cdot 10^{-4}\ \mathrm{nm}^{-2}$, $\xi_s=5$ $\mu$m and $L=10$ $\mu$m.}
\label{fig:cell}
\end{figure}

\begin{figure*}[t]
\centering
\includegraphics[width=0.95\textwidth]{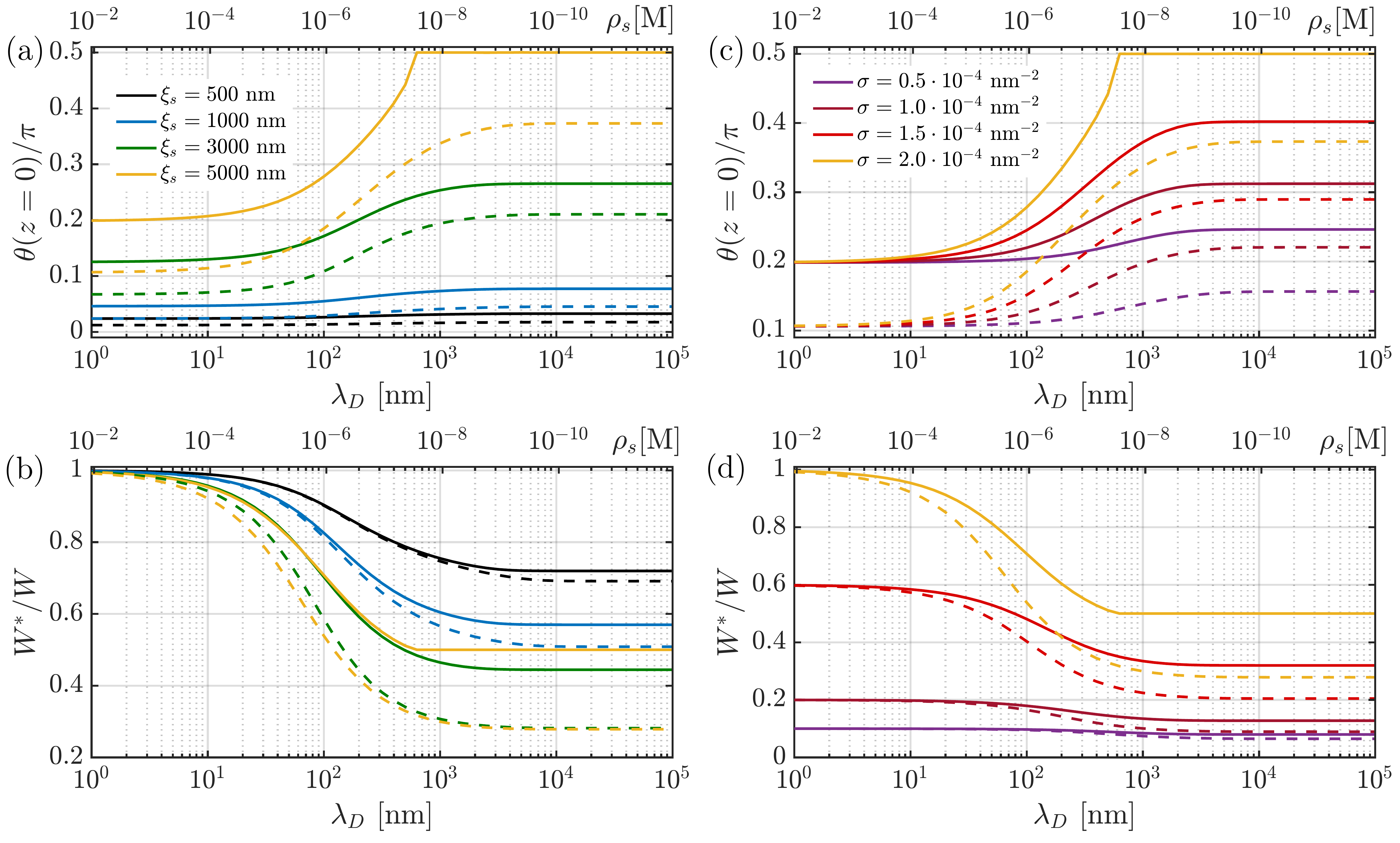}
\caption{Effective surface anchoring controlled by surface charge. Variability of the director angle $\theta$ at the bottom plate of the cell (at $z=0$) as function of Debye screening  length $\lambda_D$ or equivalently reservoir salt concentration $\rho_s$. For fixed surface charge density $\sigma=2\cdot 10^{-4}\ \mathrm{nm}^{-2}$ and varying surface extrapolation length $\xi_s$, we show (a) the surface angle and (b) the effective anchoring strength $W^*$ with respect to the bare anchoring strength $W$. The full lines are for system size $L=1$ $\mu$m, whereas the dashed lines are for $L=2$ $\mu$m. In (c) we show $\theta(0)$ for the same parameters but for fixed $\xi_s=5$ $\mu$m and varying $\sigma$ and in (d) for the same curves as in (c) the dependency of $W^*/W$ on salt concentration is highlighted.}
\label{fig:surfangle}
\end{figure*}

Electric (surface) charge introduced onto the bottom surface of the nematic cell ($\sigma\neq 0$) causes the development of an inhomogeneous electrostatic potential, and in turn the reorientation of the nematic director, as shown in Fig. \ref{fig:cell}. The electric potential close to the charged surface (at $z=0$) is large, and decreases with separation from the charged surface (increasing $z$); similarly, the net charge density in the diffuse screening cloud,
\begin{equation}
\rho_e(z)=\rho_+(z)-\rho_-(z)=-2\rho_s\sinh[\phi(z)]
\label{eq:rhoe}
\end{equation}
(see inset in Fig.~\ref{fig:cell}(a)) shows that the charge in the screening cloud decreases with $z$.  In this geometry, this is not much different as it would occur in isotropic solvents.

The resulting director profiles are shown in Fig. \ref{fig:cell}(b), where the effect of increasing surface charge $\sigma$ is shown. When $\sigma=0$ (black line in Fig. \ref{fig:cell}), the director profile is linear,
\begin{equation}
\theta(z)=\sin[2\theta(0)]\frac{z}{2\xi_s}+\theta(0),
\label{eq:linear}
\end{equation}
with $\theta(0)$ determined from the transcendental equation $\pi/2=\sin[2\theta(0)]L/(2\xi_s)+\theta(0)$. For $\sigma\neq 0$ and since $\Delta\epsilon>0$, the nematic director wants to align parallel to the electric field ($\theta=\pi/2$) due to the free-energy contribution $\sim-\Delta\epsilon({\bf n}\cdot{\bf E})^2$, with ${\bf E}$ the electric field, whereas the bottom plate has an easy axis that wants to align the nematic director perpendicular to this electric field ($\theta=0$). These two opposing effects give rise to the director profiles in Fig. \ref{fig:cell}(b), where high charge densities push the director profiles towards the constant profile $\theta(z)=\pi/2$ when electrostatic effects are much larger than the anchoring effects.

For sufficiently large $z$, the director profile shows a generally linear behaviour even for $\sigma\neq 0$, whereas the non-linear dependence on the scale of the profiles seems to be negligible. However, the non-linear part has a first derivative that cannot be necessarily neglected, and this is what mathematically drives the salt- and charge-induced anchoring transition via the boundary condition Eq. \eqref{eq:anchor}. 

In Fig. \ref{fig:surfangle} we investigate the effects of the bulk ion concentration $\rho_s$ (or equivalently $\lambda_D$), on the director at the bottom plate $\theta(z=0)$. We show the dependency of the surface angle for various system sizes $L$ (dashed lines are taken at $L=20$ $\mu$m, while full lines are taken at $L=10$ $\mu$m), extrapolation lengths $\xi_s$ and charge density $\sigma$. For small $\xi_s$ compared to the system size, the anchoring strength $W$ is large, and this means that the value of the surface angle is always close to the easy-axis value ($\theta=0$), as can be seen in Fig. \ref{fig:surfangle}(a), e.g. black lines. Conversely, if $\xi_s/L$ is large, electrostatics dominates and $\theta(0)=\pi/2$ in a large parameter regime, see full yellow line in Fig. \ref{fig:surfangle}(a). If $\xi_s$ is, however, somewhere in between these two extremes, electrostatics can compete with the anchoring, and $\theta(0)$ lies between $\theta=0$ and $\theta=\pi/2$. What is even more striking is that the surface angle is quite strongly bulk ion (salt) concentration dependent, since tuning $\lambda_D$ changes the orientation of the director on the bottom plate. 

For the surface angle dependence, we see two asymptotes, one for low Debye screening length $\lambda_D$ and one for high $\lambda_D$. For low $\lambda_D$, or equivalently high salt concentration $\rho_s$, the bottom charged plate is highly screened by the ions and $\theta(0)$ coincides with the value as if there is no surface charge on the bottom plate, see Eq. \eqref{eq:linear}. The value of the asymptote depends on the ratio $\xi_s/L$ only. The low-screening asymptote, that occurs at low salt concentration and consequently large screening length, is most relevant when the screening length becomes of the order of the system size $L$. In this case the effect of the electric field is largest and $\theta(z=0)$ becomes closer to $\pi/2$, and even $\pi/2$ when the anchoring is sufficiently weak, see, for example, the full yellow line in Fig. \ref{fig:surfangle}(a)-(b). Note, that while the non-electrostatic anchoring is weak, the homeotropic electrostatic anchoring is strong for these parameters, allowing for the control of the surface angle by the salt concentration.

The almost linear profiles (see Fig. \ref{fig:cell}(b)) allow us to define an effective anchoring strength $W^*$ in this geometry, given by the competition of the actual surface anchoring and the effect of (screened) surface charge. By mapping the case of non-vanishing $\sigma$ and $\rho_s$ on the linear profile of Eq. \eqref{eq:linear}, we find assuming that charge and salt renormalise the anchoring and not the easy-axis value, that

\begin{equation}
\frac{W^*}{W}=\frac{2[\frac{\pi}{2}-\theta(0)]}{\sin[2\theta(0)]}\frac{\xi_s}{L}.
\end{equation}
The ratio $W^*/W$ for varying $\xi_s$ is shown in Fig. \ref{fig:surfangle}(b). In line with Fig. \ref{fig:surfangle}(a) we see that an increase in $\theta(0)$, in other words, more towards homeotropic, results in a lowering of $W^*$ such that the planar anchoring is less favoured. When the bottom plate will have homeotropic alignment $\theta(0)=\pi/2$, we have that $W^*\rightarrow K/L$. For a larger system size (compare dashed with full lines), there is a larger variability of $W^*/W$ as function of $\lambda_D$, but this does not necessarily mean that the surface angle will have larger values, because $\theta(0)$ is also determined by the value of $L$.

Similar nematic behaviour  arises  upon varying the surface charge density $\sigma$, as shown in Fig. \ref{fig:surfangle}(c)-(d) for the same two different system sizes as in Figs. \ref{fig:surfangle}(a)-(b) (full and dashed lines). Again two asymptotes for the surface angle emerge as functions of the Debye screening length, where the high-screening asymptote is independent of the value of $\sigma$, whereas increasing $\sigma$ does change the value of the low-screening asymptote. In Fig. \ref{fig:surfangle}(d) we see the equivalent formulation in terms of $W^*/W$.

Nematic orientational ordering at the surface directly affects also the electric field at the surface. Indeed, for all regimes of surface nematic director shown in Fig. \ref{fig:surfangle}, the
normal component of the electric field ${\bf E}$ at the surface, given as
\begin{equation}
E_z(0)=\frac{4\pi\lambda_B\bar{\epsilon}k_BT\sigma}{q_e[\epsilon_\perp+\Delta\epsilon\sin^2\theta(0)]},
\end{equation}
decreases for increasing $\lambda_D$, since $\theta(0)$ increases. This could be considered as  counterintuitive since one would expect that a larger magnitude of the electric field at the bottom plate's surface would also reorient the director more parallel to the electric field, but in Fig. \ref{fig:surfangle} we see the opposite. This shows that the value of the surface angle depends more on the ability of the electric field to distort the nematic from the zero-$\sigma$ case sufficiently far from the plate, than the ability of the electric field to distort the nematic close to the bottom plate. This situation is reminiscent of the Fréedericksz transition \cite{Frederiks:1933}, where the electric coherence length that quantifies the ability of the electric field to overcome the elastic interactions, should be sufficiently large in order to have a transition in the nematic cell \cite{deGennes:1993}.

\begin{figure}[t]
\centering
\includegraphics[width=0.5\textwidth]{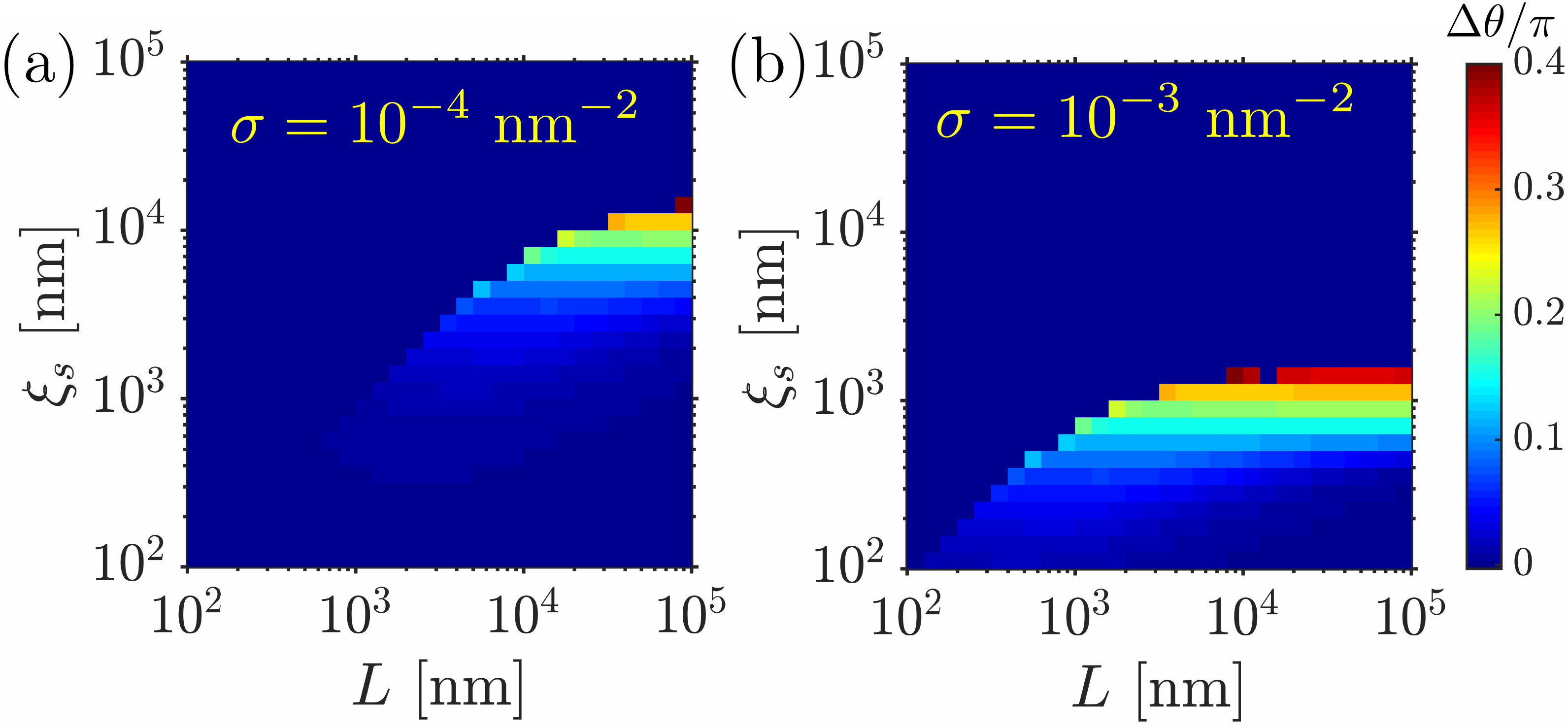}
\caption{Tunability of the surface-angle difference $\Delta\theta$ between screening lengths of 1 nm and the low-screening regime ($\lambda_D>L$) as function of system size $L$ and surface extrapolation length $\xi_s$, for two surface charge densities $\sigma$.}
\label{fig:tune}
\end{figure}

This leads us to the natural question: ``For what parameter values of $\xi_s$, $L$ and $\sigma$ is the surface angle the most tuneable by varying the salt concentration?". We attempt to answer this question by calculating the value of $\theta(0)$ at the low-screening asymptote and to compare it with the value at $\lambda_D=1$ nm, and consider their difference $\Delta\theta=\theta(z=0,\lambda_D\rightarrow\infty)-\theta(z=0,\lambda_D=1\ \mathrm{nm})$.
We do not use the high-screening asymptote in this comparison since for some set of parameters, we find $\theta(0)=\pi/2$ in the range of $\lambda_D$ between 1 nm and $10L$. In these cases, the high-screening asymptote is only reached at low, unphysical values of $\lambda_D$, which are similar or smaller than the size of the molecular building blocks, and then our continuum theory breaks down. 

We plot $\Delta\theta$ in Fig. \ref{fig:tune} for varying $\xi_s$ and $L$ for two values of $\sigma$. We see that the surface angle is most tunable for $\xi_s<L$ (for $\xi_s>L$ the director would align in a completely homeotropic manner), and that the specific values of $\xi_s$ and $L$ for which it is tuneable depends on $\sigma$: for low $\sigma$ this occurs at higher values of $\xi_s$ and $L$ than for high $\sigma$, for which the tendency to reorient towards a homeotropic alignment is larger. We conclude that for surface charges where the electrostatic homeotropic anchoring can compete with the non-electrostatic planar anchoring, we find possible tunability of the total anchoring by the salt concentration.

Finally, in this section, we assumed that the nematic dielectric anisotropy is positive, $\Delta\epsilon>0$, so the director tends to lie parallel to the electric field. In the case where $\Delta\epsilon<0$, the dielectric coupling between nematic director and electric field would prefer the director to align perpendicular to ${\bf E}$, and in this case $\theta(0)$ would decrease with the Debye screening length $\lambda_D$. Such behaviour, for example, could be realised by using nematic materials like MBBA \cite{Meyerhofer:1975}. 

%As a comment to the above analysis, note, that results are shown also for $\lambda_D\sim 0.1\ \mathrm{nm}$, which are screening lengths comparable to the size of the actual typical ion, where continuum approach used in this work can become insufficient.

\section{Charge-regulation effects on the anchoring-strength tunability}
\label{sec:CR}

\begin{figure}[t]
\centering
\includegraphics[width=0.48\textwidth]{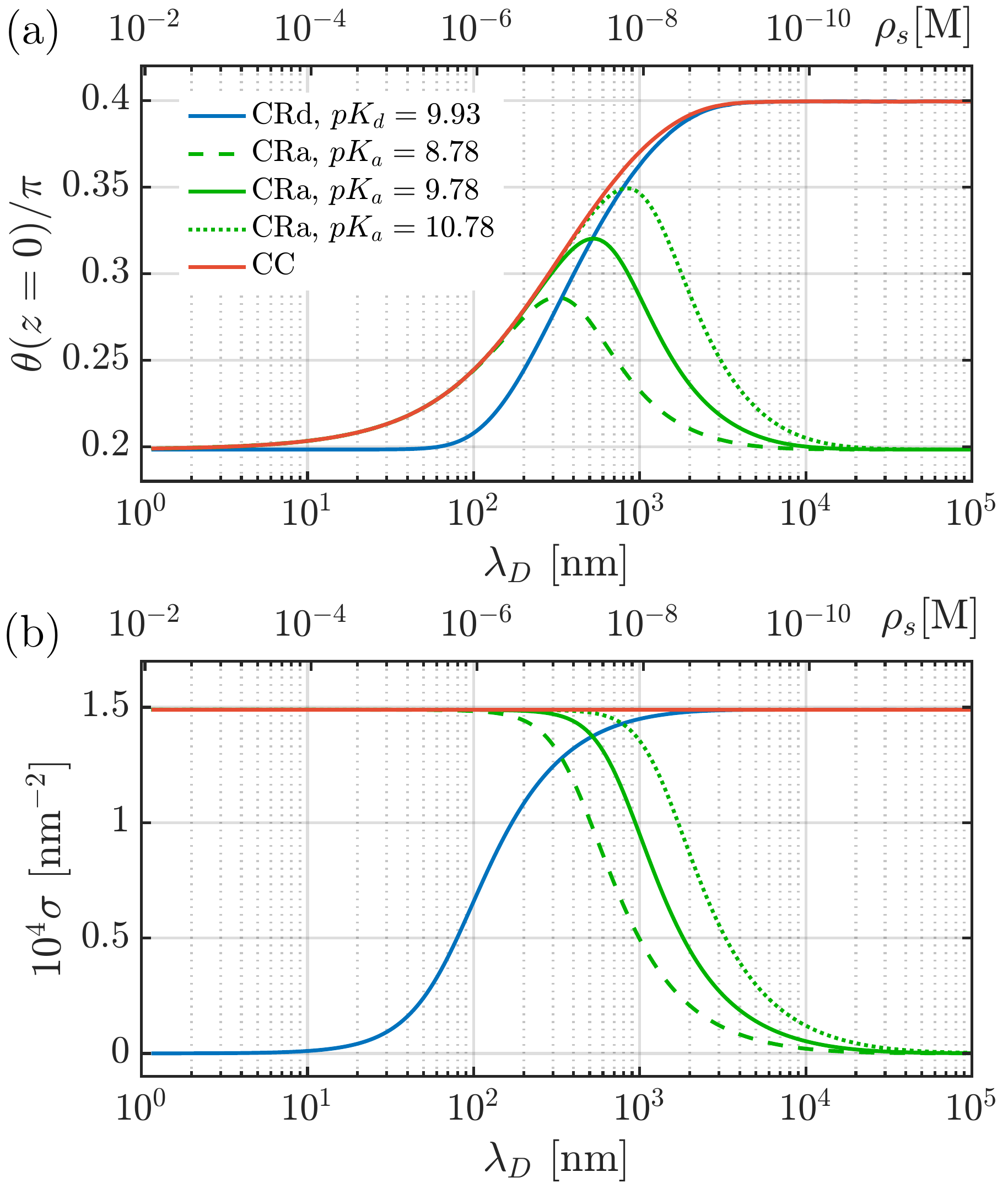}
\caption{Effect of charge regulation on (a) the surface angle $\theta(0)$ and (b) the surface charge density $\sigma$ on the bottom plate as function of the isotropic screening length $\lambda_D$ or equivalently the reservoir salt concentration $\rho_s$. The various reaction mechanisms, charging via association of a cation (CRa) or dissociation of an anion (CRd), are compared with the constant-charge case (CC) for various $pK$ values. We choose $\sigma$ in the CC case such that it matches the high-screening $\sigma$ of the CRa cases, and the low-screened asymptote of the CRd case. {\color{black} $pK$ values are chosen such that a maximum surface charge density of $1.5\times 10^{-4}$ nm$^{-2}$ is reached.} In all plots, we set surface extrapolation length $\xi_s=5\ \mu$m and system size $L=10\ \mu$m.}
\label{fig:CR}
\end{figure}

Realistic chargeable surfaces are not necessarily constant charge over the full range of salt concentrations what we have assumed in Sec. \ref{sec:chargean}. Therefore, charge regulation is of major interest in soft matter applications, from the charging behaviour of colloidal particles \cite{Lowen:2014, Linden:2015, Markovich:2016, Trefalt:2016, Bartlett:2018, Levin:2019}, to the effect on interaction potentials \cite{Everts:2017}, to phase behaviour \cite{Bedzyk:2013, Everts:2016}, but it is also important in biological context. We use the Langmuir adsorption isotherms \cite{Ninham:1971}
and consider two different systems that are characterised by the following chemical reactions,

\begin{align}
\mathrm{S}_a+\mathrm{P}^+\leftrightarrows \mathrm{S}_a\mathrm{P}^+,& \quad K_a=\frac{\vartheta_{\mathrm{S}_a}[\mathrm{P}^+]}{\vartheta_{\mathrm{S}_a\mathrm{P}^+}},& \quad  (\mathrm{CRa}), \\
\mathrm{S}_d\mathrm{N}\leftrightarrows \mathrm{S}_d^++\mathrm{N}^-,& \quad K_d=\frac{\vartheta_{\mathrm{S}_d^+}[\mathrm{N}^+]}{\vartheta_{\mathrm{S}_d\mathrm{N}}},& \quad (\mathrm{CRd}),
\end{align}
where $\mathrm{S}_a$ is a neutral surface site that acquires its charge by adsorption of a cation ($\mathrm{P}^+$), and $\mathrm{S}_d\mathrm{N}$ is a surface site that can positively charged by losing an anion ($\mathrm{N}^-$). The equilibrium constants for these reactions are given by $K_i$, $(i=a,d)$, $\vartheta_i$ denote surface coverages (fraction of sites occupied), and square brackets denote the concentration of the particular species. {\color{black} From a microscopic point of view equilibrium constants can be formulated in terms of a binding energy and a binding volume that accounts for all configurations that correspond to a chemical bond. Therefore, the $K_i$ values do not depend explicitly on surface coverages nor concentrations. Furthermore, recently it was shown within a minimal sticky hard-sphere model that the equilibrium constant of a surface group on a chargeable surface does not equal the equilibrium constant of the \emph{same} functional group in single molecules, see for details Ref. \cite{Levin:2019}.

We derive expressions for the surface charge density of both chemical reactions,
\begin{align}
\sigma=\sigma_m\left\{1+\dfrac{K_a}{\rho_s}\exp[\phi(0^+)]\right\}^{-1}, \quad (\mathrm{CRa}), \label{eq:CRa} \\ \quad
\sigma=\sigma_m\left\{1+\dfrac{\rho_s}{K_d}\exp[\phi(0^+)]\right\}^{-1}, \quad (\mathrm{CRd}) \label{eq:CRd}.
\end{align}  
Here we used that the total number of sites is conserved, and that the concentrations can be expressed as \mbox{$[\mathrm{P}^+]=\rho_s\exp[-\phi(0^+)]$} and $[\mathrm{N}^-]=\rho_s\exp[\phi(0^+)]$. Furthermore,} $\sigma_m$ is the surface number density of chargeable sites. Eqs. \eqref{eq:CRa} and \eqref{eq:CRd} will be used in boundary condition Eq. \eqref{eq:chargeboundary}, rather than choosing a fixed value of $\sigma$. We note that these expressions can also be obtained by adding an appropriate surface free energy to the total free energy \cite{Everts:2017}. These two reactions behave differently as function of salt concentration and hence of $\lambda_D$. Surfaces with reaction CRa discharge for increasing $\lambda_D$ (decreasing $\rho_s$). The maximal surface charge density $\sigma_\text{max}$ equals $\sigma_m$. In contrast, surfaces with reaction CRd acquire a higher surface charge for increasing $\lambda_D$. In this case and for our choice of parameters, we find $\sigma_\text{max}\ll\sigma_m$ since in this confined geometry a Donnan potential is generated $\phi(L)> 0$ at high $\lambda_D$.

\begin{figure*}[t]
\centering
\includegraphics[width=0.95\textwidth]{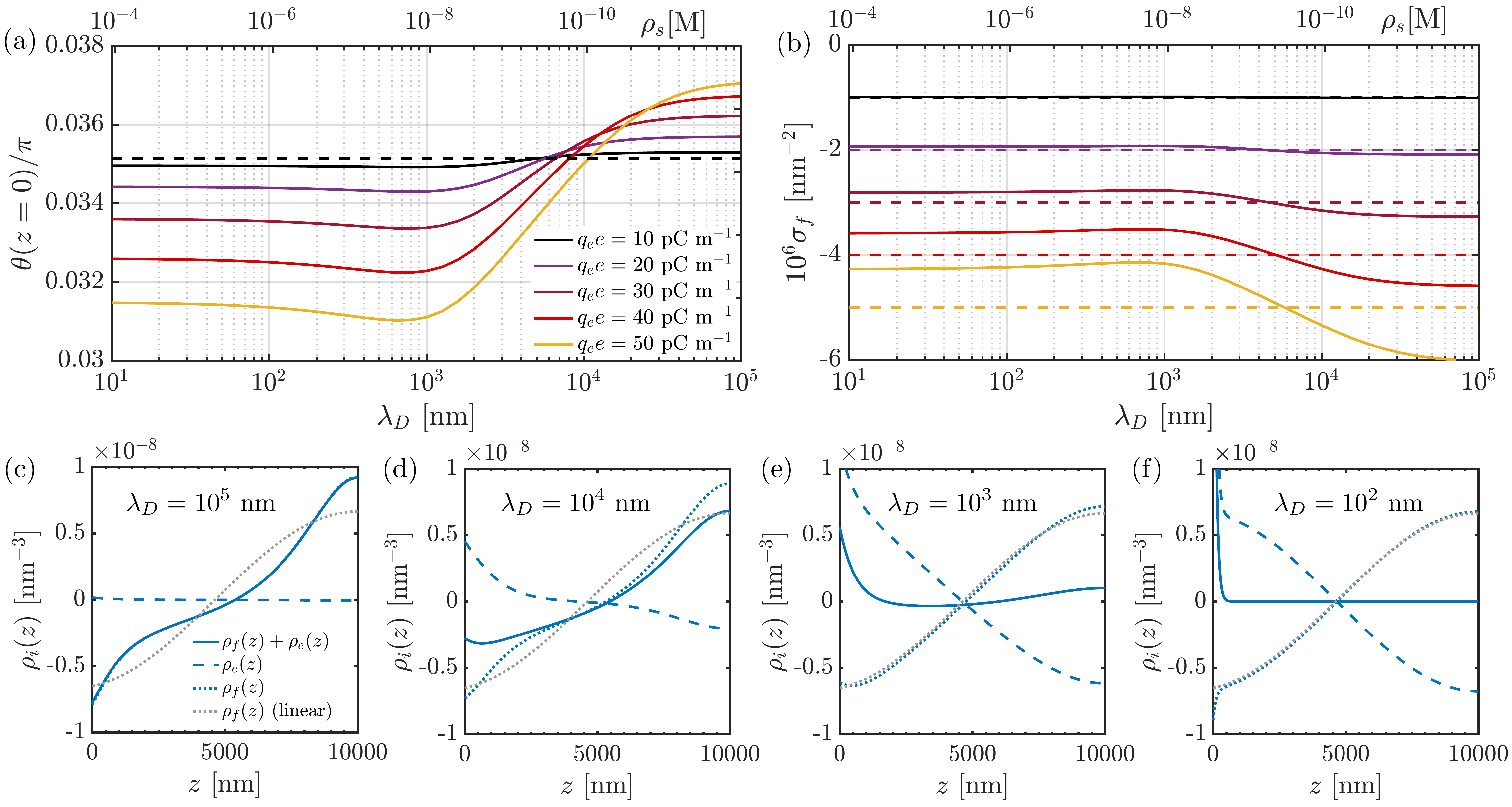}
\caption{Flexoelectricity-driven control of nematic surface ordering for uncharged surfaces for system size $L=10$ $\mu$m and surface extrapolation length $\xi_s=750$ nm. (a) The nematic director surface angle $\theta(z=0)$ as function of isotropic screening length $\lambda_D$ (or equivalently reservoir salt concentration $\rho_s$) for different flexoelectric coefficients $q_ee$ and (b) the resulting flexoelectric bound surface charge charge densities $\sigma_f$. The dashed lines are the values of $\theta(0)$ and $\sigma_f$, respectively, when a linear profile (which is achieved when $e=0$) is assumed. For the highest value $q_ee=50$ pC m$^{-1}$, we show in (c)-(f) the net ion charge density $\rho_e(z)$ and the flexoelectric polarisation charge density $\rho_f(z)$ profiles and their sum, for selected values of $\lambda_D$. The grey dotted line is $\rho_f(z)$ when the director profile would be linear.}
\label{fig:flexonosurfcharge}
\end{figure*}

In Fig. \ref{fig:CR} we compare the CRd and CRa case with the constant-charge (CC) case. We chose parameters such that the plot highlights the differences best. We choose values of $K_i$ and $\sigma_m$ such that the CR cases saturate at $\sigma_\text{max}\approx 1.5\times10^{-4}\ \mathrm{nm}^{-2}$, for CRd at high $\lambda_D$, and for CRa at low $\lambda_D$. For CRd we took $\sigma_m=1$ nm$^{-2}$, but the precise value is not so relevant as it is for the CRa case. For the CC case we took $\sigma=\sigma_\text{max}$. In Fig. \ref{fig:CR}(b) we plot $\sigma$ as function of $\lambda_D$, showing the charging/discharging differences between the CRa and CRd cases.

When compared to the CC case, we see that reaction CRd reaches its high-screening asymptote at a higher value of $\lambda_D$ than the CC case, see Fig. \ref{fig:CR}(a). The reason is that the surface charge $\sigma$ is lower at these salt concentrations, see Fig. \ref{fig:CR}(b), because of charge regulation. The low-screening asymptote coincides with that of the CC case by construction. Hence, the tunability range at the same $\sigma$ for CRd is smaller than for CC and the value of $K_d$ tunes the value of $\sigma_\mathrm{max}$.

On the other hand, for this set of parameters we see that CRa gives a non-monotonous behaviour for $\theta(0)$ as function of $\lambda_D$. This non-monotonous behaviour emerges because of two opposing effects. Firstly, $\sigma$ decreases for increasing $\lambda_D$, and secondly, when $\lambda_D$ increases the charge becomes more important to reorient the director towards $\pi/2$. When $\sigma$ reaches $\sigma_\text{max}$ the curve follows the CC curve with the same value of $\sigma$ as the $\sigma_\text{max}$ of the CRd case. The value of $\sigma_\text{max}$ is set by $\sigma_m$ rather than the value of $K_a$ (in contrast with the CRd case). $K_a$ instead sets the value of $\lambda_D$ at which the plate starts to discharge for increasing $\lambda_D$. This means that we can shift the maximum of $\theta(0)$ as function of $\lambda_D$ by tuning $K_a$ (compare the green lines in Fig. \ref{fig:CR}(a)). For 5CB the most interesting range is to have $\sigma$ tunable in the 100 nm-1 $\mu$m regime, these are Debye screening lengths that are still reachable by doping 5CB, and it is easier to dope than to deionise a liquid crystal.

Finally, we note that the two reaction mechanisms CRa and CRd give so remarkably different anchoring behaviour that by measuring the director orientation one can learn something of the reaction mechanism on how a general surface becomes charged. This is a valuable observation because for some surfaces the charging process is not known in detail and one has to hypothesise which chemical reactions are relevant for the surface charging. For example, our results suggest that the planar to homeotropic transition observed in Ref. \cite{Bungabong:2010} is of the CRa type.

\section{Flexoelectricity-driven anchoring control}
\label{sec:flexo}

Flexoelectricity is another electrostatic mechanism for controlling the effective surface anchoring, as an alternative or as an addition to the surface alignment control by the surface charge. Because of the geometric constraints in the nematic cell under consideration, we can tune the strength of the flexoelectricity by a single parameter $e=e_1+e_3$, allowing us to access different material regimes. In contrast to just having a surface charge, flexoelectricity results in a bound space-charge density to which the ions can couple. 

Flexoelectricity-driven  control of surface ordering is shown in Fig. \ref{fig:flexonosurfcharge}.
In Fig. \ref{fig:flexonosurfcharge}(a), the nematic director surface angle $\theta(0)$ is shown for different screening lengths $\lambda_D$, and we see that upon increasing $\lambda_D$, initially $\theta(0)$ decreases, and then increases when $\lambda_D$ exceeds $\sim 10^3$ nm. This effect is rather weak even for high flexoelectric coefficients, with the change in the director angle of the order of $1^{\circ}$, and is also independent of the sign of $e$, because the EL equations have an internal symmetry $\phi\rightarrow-\phi$ for $\sigma=0$. Therefore, this phenomenon is only of theoretical interest. The dashed black line in Fig. \ref{fig:flexonosurfcharge}(a) shows the case where $\sigma=0$ and $e=0$, and corresponds to the linear profile as given by Eq. \eqref{eq:linear}.

In order to understand this small, non-monotonous, non-linear behaviour upon varying $\lambda_D$, we consider the flexoelectric surface bound charge density on the bottom plate, given by $\sigma_f=-{\bf P}_f\cdot\hat{{\boldsymbol{\nu}}}$, with $\hat{\boldsymbol{\nu}}$ an outward-pointing unit normal, such that
\begin{equation}
\sigma_f=-\frac{e}{2}\sin[2\theta(0)]\theta'(0).
\label{eq:sigmaf}
\end{equation}
In Fig. \ref{fig:flexonosurfcharge}(b), we plot $\sigma_f$ for various values of $e$. The dashed lines indicate the situation where we take the linear profile of Eq. \eqref{eq:linear} and put it in Eq. \eqref{eq:sigmaf}. This can be interpreted as if the presence of flexoelectricity would not perturb the director profile, but still generate a flexoelectric polarisation since there is a director gradient. We see that $\sigma_f$ follows the same trend as $\theta(0)$ as function of salt concentration: when $\sigma_f$ is high (more negative) $\theta(0)$ is also larger and vice versa.

\begin{figure*}[t]
\centering
\includegraphics[width=0.9\textwidth]{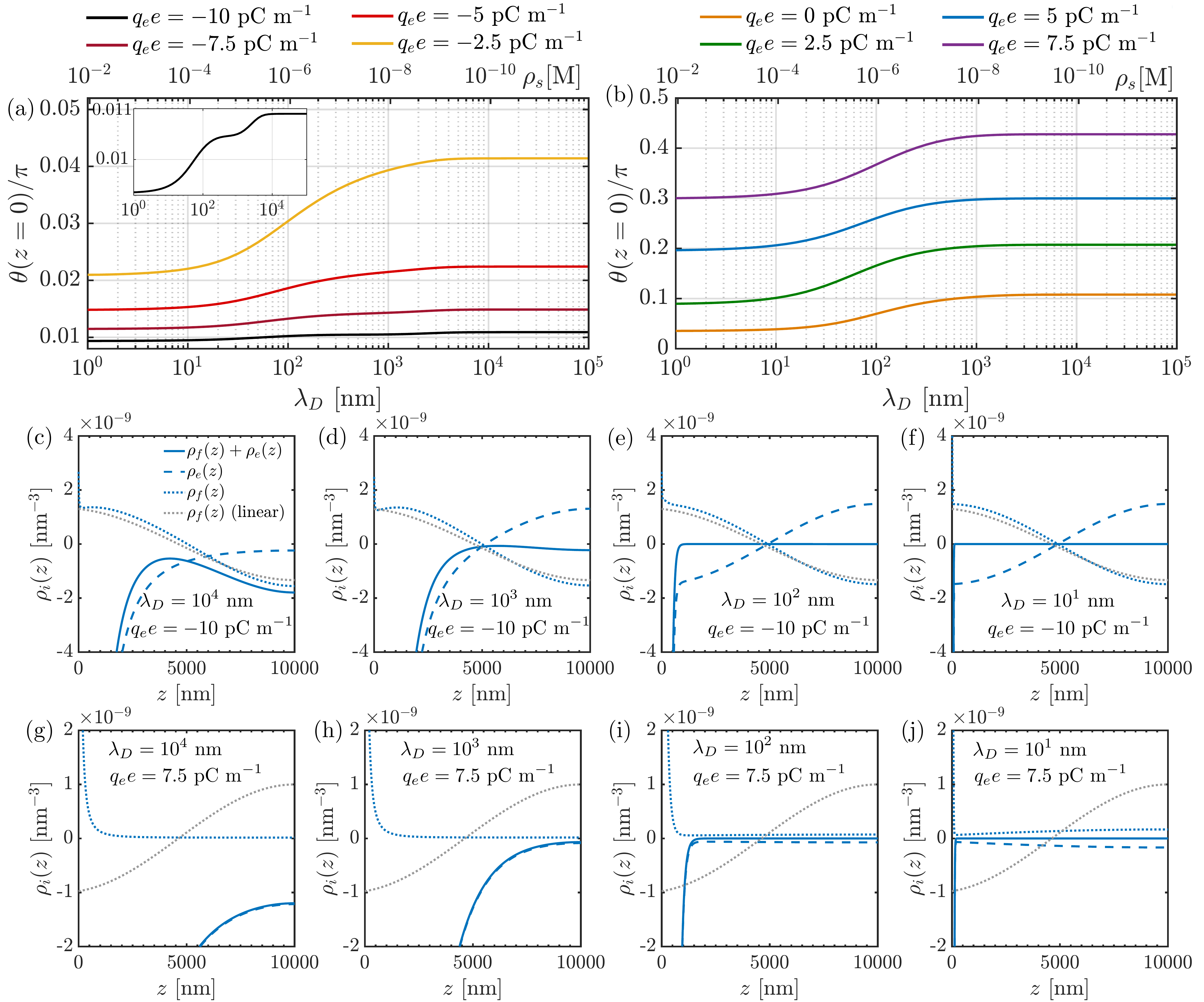}
\caption{Flexoelectricity-driven control of nematic surface ordering for a charged bottom cell surface ($\sigma=5\cdot10^{-4}$ nm$^{-2}$) for system size $L=10$ $\mu$m and surface extrapolation length $\xi_s=750$ nm. (a) The nematic director surface angle $\theta(z=0)$ as function of isotropic screening length $\lambda_D$ (or equivalently reservoir salt concentration $\rho_s$) for (a) negative flexoelectric coefficients $q_ee$ and (b) positive $q_ee$. The inset in (a) shows a zoomed-in version of $\theta(0)$ at $q_e e=-10$ pC m$^{-1}$. For this value of $q_ee$, we show in (c)-(f) the net ion charge density $\rho_e(z)$ and the flexoelectric polarisation charge density $\rho_f(z)$ profiles and their sum, for selected values of $\lambda_D$. The grey dotted line is $\rho_f(z)$ when the director profile would be linear. In (g)-(j) we show the same as in (c)-(f), but for $q_e e =7.5$ pC m$^{-1}$. }
\label{fig:flexopluscharge}
\end{figure*}

What is surprising is that at the no-salt limit (high $\lambda_D$) $\sigma_f$ is higher than the unperturbed state, the system prefers to have a surface flexoelectric polarisation. This can be rationalised when we also consider the bulk, where a flexoelectric polarisation (volume) charge density $\rho_f=-\nabla\cdot {\bf P}_f$ is generated, given by
\begin{equation}
\rho_f(z)=-\frac{e}{2}\left\{2\cos[2\theta(z)][\theta'(z)]^2+\sin[2\theta(z)]\theta''(z)\right\}.
\end{equation}
In the low-screening limit we plot $\rho_f(z)$ in Fig. \ref{fig:flexonosurfcharge}(c), and in dotted grey we plot the situation when the director profile would not be perturbed. The salt concentration is negligible. We see that close to the bottom plate, a higher (more negative) $\rho_f$ is found compared to the unperturbed state (compare full blue with dotted grey line), and the same happens close to the top plate, albeit the top plate has no surface flexoelectric polarisation due to the strong homeotropic boundary conditions. The cost of having a flexoelectric polarisation charge density close to the top and bottom plate is compensated by the fact that in the middle of the cell $\rho_f(z)$ is reduced compared to the unperturbed state. The region where it is reduced is larger than the region where $\rho_f(z)$ is enhanced. Reducing $\rho_f(z)$ in the middle of the cell at the expense of increasing it close to the plates cannot happen indefinitely because this would also cost more elastic energy.

When salt is added we see that a diffuse screening cloud is formed with charge density $\rho_e(z)$ as given by Eq. \eqref{eq:rhoe}, see the dashed lines in Fig. \ref{fig:flexonosurfcharge}(d), that is positively charged close to the bottom plate and switches sign roughly in the other half of the nematic cell. This diffuse ion cloud effectively screens the flexoelectric polarisation charge (blue dotted line), and is doing this the most effectively near the bottom and the top plate (compare blue full line with dotted blue line). This causes $\sigma_f$ to reduce (becoming less negative), and this ultimately also enhances $\rho_f(z)$ in the middle of the plate, whereas $\rho_f(z)$ close to the plates becomes close to the unperturbed state (compare dotted blue with dotted grey in Fig. \ref{fig:flexonosurfcharge}(e)). The bulk $\rho_f(z)$ is then almost effectively screened, until it is completely screened as we can see in the full blue line in Fig. \ref{fig:flexonosurfcharge}(f)). However, there is not enough salt to completely screen the surface, as is seen from $\rho_f(z)+\rho_e(z)>0$ close to the bottom plate. This allows $\sigma_f$ to become more negative again upon adding salt (decreasing $\lambda_D$).

When there is a free surface charge $\sigma\neq0$ the symmetry $e\rightarrow-e$ is lost, and we have to consider $e<0$ and $e>0$ separately. In Fig. \ref{fig:flexopluscharge}(a) we show the effects of ions on $\theta(0)$ in a nematic liquid crystal with negative flexoelectric coefficient.  First of all, we see that a more negative $e$ destabilises the homeotropically aligned surface state for $\sigma>0$ and $\Delta\epsilon>0$. This is in accordance with the surface free energy analysis of Ref. \cite{Petrov:1993}. Secondly, for varying salt concentration, we observe that a more negative $e$ reduces the high-screening asymptote and reduces the tunability range of the surface angle. However, the profile of the most negative $e$ shows the most interesting characteristics for $\theta(0)$, revealing three plateaus as function of $\lambda_D$ of theoretical interest. For all values of $\lambda_D$ the value of $\rho_f(z)$ is similar to the unperturbed state (Fig. \ref{fig:flexopluscharge}(c)-(f)) compared to the case where $\sigma=0$ where this only occurs for sufficiently high $\rho_s$ (Figs. \ref{fig:flexonosurfcharge}(e)-(f)). At high $\lambda_D$ there is not enough salt available to screen $\rho_f(z)$, so the available salt screens the free surface charge on the bottom plate while only partially screening $\rho_f(z)$ in the bulk, see Fig. \ref{fig:flexopluscharge}(c), compare full blue line with dotted blue line. Upon lowering $\lambda_D$, the ions are able to screen $\rho_f(z)$ in the bulk. When this occurs, increasing $\rho_s$ only leads to more screening of $\sigma$ which lowers the surface angle, but because of the negative value of $\sigma_f$, the value of the low-screening asymptote is lower than that of the case where $e=0$.

For $e>0$, we see that increasing $e$ results in an overall higher value of the surface angle, again in accordance with the analysis of Ref. \cite{Petrov:1993}, and a larger tunability of this quantity, see Fig. \ref{fig:flexopluscharge}(b), compared to the unperturbed state. This enhancement towards homeotropic anchoring is accompanied by a $\rho_f(z)$ that is not so sensitive to the value of $\lambda_D$ unless $\lambda_D$ is very small, see Fig. \ref{fig:flexopluscharge}(g)-(i), and it is very different than the $\rho_f(z)$ of the unperturbed state, with even a different sign close to the bottom plate. For this configuration it seems to be more energetically favourable to have no bulk flexoelectric polarisation, only close to the bottom plate, and having a positive $e$ allows for this. Increasing $\rho_s$ (lowering $\lambda_D$) enhances the screening of the bottom plate and hence reduces the surface angle. There is also a flexoelectric surface charge density, which in the full range of $\lambda_D$ is of the order of $-10^{-6}$ nm$^{-2}$ and is therefore too low to compete with the free surface charge already present on the plate. These examples highlight the complex interplay when flexoelectricity is added, because it couples in a nontrivial way with (ion) electrostatics and the director profile.
\section{Conclusions and outlook}
\label{sec:con}

In this paper, we have investigated the effects of surface charge, ion (salt) bulk concentration and flexoelectricity on the effective alignment of a nematic fluid at a surface. We have investigated the specific case of a plate that enforces planar anchoring that competes with the homeotropic alignment due to the effects of surface charge. We have found that electrostatic screening can tune the electrostatic part of the alignment, in other words, by adding salt one can effectively tune the anchoring strength of a surface. The tunability is largest when the electrostatic anchoring is large enough to compete with the non-electrostatic anchoring. We have highlighted the role of the various parameters, such as bare anchoring strength, system size, surface charge and salt concentration, and their interplay.

Besides discussing surfaces that have a constant surface charge independent of the salt concentration, we have also considered charge-regulating surfaces that acquire their surface charge by ad- or desorption of ions. This directly influences the tunability of the orientation of the director at the surface, and in particular, the dependence on the salt concentration is different if the charging occurs via a dissociative or an associative process. Namely, for an associative charging, surfaces discharge upon \emph{decreasing} the salt concentration, while dissociative charging leads to surfaces discharging with \emph{increasing} the salt concentration. This shows that measuring the director tunability can reveal information on the surface chemistry and the charging properties of external surfaces.

Furthermore, we have considered flexoelectric effects on the alignment of the director at an external surface. We found that even in the absence of surface charges the director is tunable by the salt concentration; however, the variability of the director orientation is very small. When a free surface charge is added, we found that negative flexoelectric coefficients reduce the tunability of the director orientation at the surface, and makes the orientation more planar. In contrast, positive flexoelectric coefficients enhance the tunability and promote more homeotropic alignment for the full range of salt concentrations.

Our findings have clear relevance in various fluids with nematic orientational order, most evidently in nematic fluids, as charges and ionic impurities are regularly present in such materials. We show the surface ordering in the context of an idealised flat-plate geometry, where only specific elastic and flexoelectric modes can be excited; however, more complex systems and geometries would respond  in a similar manner, offering very interesting advanced routes for the microscopic design of even more complex electric charge and electric potential profiles. Nevertheless, it would be appealing to also consider theoretically the inclusion of more elastic and flexoelectric modes, as well as going beyond the treatment of the effective Rapini-Papoular surface free energy that is used in this work, e.g. by using simulations. Finally, we envisage that the geometries with colloidal particles in nematics are interesting because ions could influence the anchoring strength depending on the exact configuration of particles. We hope that our paper will stimulate experimental and theoretical work on ion-doped liquid crystals, rather than viewing ions as an impurity, with a special emphasis on controlling the salt concentration in the nematic host.

\section*{\normalsize{Acknowledgements}}
J. C. E. acknowledges financial support from the European Union's Horizon 2020 programme under the Marie Skłodowska-Curie grant agreement No. 795377. M. R. acknowledges financial support from the Slovenian Research Agency ARRS under contracts P1-0099, L1-8135 and J1-1697. The authors acknowledge I. Smalyukh and S. \v Copar for fruitful discussions. Finally, the authors would like to thank the Isaac Newton Institute for Mathematical Sciences for support and hospitality during the programme [The Mathematical Design of New Materials] when part of this work on this paper was undertaken. This work was supported by: EPSRC grant number EP/R014604/1.

\bibliographystyle{apsrev4-1} % Tell bibtex which bibliography style to use
\bibliography{literature1} % Tell bibtex which .bib file to use (this one is some example file in TexLive's file tree)

\end{document}